\documentclass[%
 reprint,
 amsmath,amssymb,
 aps,
]{revtex4-2}

\usepackage{graphicx}
\usepackage{dcolumn}
\usepackage{bm}
\usepackage{hyperref}
\hypersetup{colorlinks=true,allcolors=[RGB]{40,195,190}}

\begin{document}

\preprint{APS/123-QED}

\title{Photon radiation induced by rescattering in strong-interacting medium with a magnetic field}

\author{Yue Zhang}
\email{yue.zhang@mails.ccnu.edu.cn}
\affiliation{Institute of Particle Physics and Key Laboratory of Quark and Lepton Physics (MOE), Central China Normal University, Wuhan 430079, China}
\author{Han-Zhong Zhang}
\email{zhanghz@mail.ccnu.edu.cn}
\affiliation{Institute of Particle Physics and Key Laboratory of Quark and Lepton Physics (MOE), Central China Normal University, Wuhan 430079, China}

\date{\today}

\begin{abstract}

The photon radiation induced by rescattering in a magnetized medium is investigated in relativistic heavy-ion collisions. Within the high-energy limit, the photon emission rate and the associated electromagnetic energy loss are derived using the Gyulassy-Levai-Vitev formalism at first order in opacity, for a quark jet propagating a quark-gluon plasma under a background magnetic field. Quantitative analysis shows a slight suppression of the overall photon radiation over a broad range of jet energies in this process. This reduction in photon yield consequently leads to a moderate decrease in the electromagnetic energy loss of the jet. Our results contribute to a better understanding of the electromagnetic properties of strongly interacting matter in high-energy nucleus-nucleus collisions and motivate experimental comparison of photon yields from quark-gluon plasma with similar properties but distinct magnetic field strengths.

\end{abstract}
\maketitle
\section{\label{sect:1}Introduction}

A large number of theoretical analysis \cite{Gross:1980br,Gyulassy:1990ye,Gyulassy:2004zy} and experimental results \cite{Shuryak:2004cy,BRAHMS:2004adc,Braun-Munzinger:2015hba} have shown that high-temperature and high-density deconfined Quark-Gluon Plasma (QGP) might be produced in the heavy-ion collisions under the energy of the Large Hadron Collider (LHC) at European Organization for Nuclear Research and the Relativistic Heavy Ion Collider (RHIC) at Brookhaven National Laboratory. Many types of probes have been developed to investigate the properties of nuclear matter in heavy-ion collisions, among which electromagnetic probes---real and virtual photons---offer distinct advantages \cite{Alam:1996fd}. Electromagnetic signals are produced throughout all stages of collisions, from the initial hard scattering to the subsequent evolution of the nuclear matter. Unlike strongly interacting particles, photons, once produced, scarcely interact with other particles in the medium, thereby endowing electromagnetic probes with mean free paths much longer than the size of the fireball created in heavy-ion collisions \cite{Cassing:1999es}. Because electromagnetic probes possess these properties, they can provide direct information about the microscopic processes and their surrounding environment and serve as valuable probes for uncovering the unique characteristics of strongly interacting system \cite{Linnyk:2015rco}. Extensive theoretical and experimental studies of electromagnetic probes have been carried out over the past decades \cite{Linnyk:2012pu,Wu:2024vyc,PHENIX:2009gyd,ALICE:2022hvk}.

In high-energy nucleus-nucleus (AA) collisions, real photons detected in experiments originate from multiple sources, including Compton scattering and quark-antiquark annihilation in initial parton hard scatterings, jet fragmentation, medium-induced jet bremsstrahlung, QGP thermal radiation, hadronic decay and so on \cite{Linnyk:2015rco,David:2019wpt}. Those produced from all sources except hadronic decays are commonly referred to as direct photons \cite{David:2019wpt}.
These photons, emitted via diverse mechanisms throughout the entire space–time evolution of hot and dense nuclear medium, provide a unique and powerful means to probe both the initial conditions and the subsequent evolution of the QGP \cite{David:2019wpt,Gale:2018vuh,Shen:2023aeg}.
Theoretical studies have analyzed the yields and compositions of direct photons ~\cite{Turbide:2007mi,Linnyk:2015tha,Xie:2020zdb} and compared them with experimental measurements \cite{Buesching:2005hpv,Isobe:2007ku,WA98:2000vxl,PHENIX:2008uif,PHENIX:2014nkk}.
However, the diversity and complexity of direct photon production processes pose significant challenges in precisely estimating and separating the contributions from different sources. As a result, several issues related to photon production remain unresolved \cite{Turbide:2005bz,Shen:2016odt}.
This work focuses on detailed analysis of medium-induced jet bremsstrahlung photon in AA collisions, aiming to contribute to completing the puzzle of photon production and addressing remaining open questions.

Medium-induced photon radiation is also closely related to hard probes, namely the jet and its associated energy loss mechanisms. When an energetic jet propagates through the QGP, it interacts strongly with the medium, inducing gluon radiation and resulting in substantial energy loss \cite{Gyulassy:1990ye}. This phenomenon is known as jet quenching effect \cite{Qin:2015srf}. Within the framework of quantum chromodynamics (QCD), jet quenching effect has been investigated in detail, and it has been found to perfectly explain experimental phenomena such as single hadron, dihadron suppressions and di-jet transverse momentum imbalance \cite{Wang:1992qdg,Han:2022zxn,Gao:2023lhs}. 
For quark jet, multiple scattering can induce not only gluon radiation but also photon emission \cite{Zakharov:2004bi,Zhang:2010hiv,Zhang:2016avg}. Research on medium-induced photon radiation and electromagnetic energy loss of jet in heavy-ion collisions therefore provides an additional way for exploring the formation and properties of the QGP.

In non-central AA collision experiments, the relative motion of the two charged, ultrarelativistic beams generates an intense background magnetic field \cite{Hattori:2023egw,Siddique:2025tzd,Shen:2025unr}. The component of this field perpendicular to the reaction plane is estimated to reach strengths of approximately $eB\sim 5m_{\pi}^2 \sim 10^{9}\mathrm{G}$ at RHIC and $eB\sim 50m_{\pi}^2 \sim 10^{10}\mathrm{G}$ at LHC \cite{Kharzeev:2007jp,Skokov:2009qp,Voronyuk:2011jd,Deng:2012pc,Tuchin:2013ie}.
Unlike in vacuum, where such a field would vanish rapidly, the finite electrical conductivity of the QGP significantly delays its decay \cite{Tuchin:2013apa,McLerran:2013hla,Li:2016tel}, potentially leading to non-negligible effects on the dynamics of quark jets traversing the QCD medium \cite{Avancini:2011zz,Zhang:2024htn}.

To study the multiple scattering of a hard jet in a thermomagnetic medium, the Gyulassy-Wang (GW) model is employed to describe the interaction between the jet and target partons via a static Yukawa potential \cite{Gyulassy:1993hr}. Since the primary interest lies in relatively low momentum transfer scattering, the spin of the partons can be neglected. The quark jet parton is described by a charged scalar propagator in this model, which serves as a suitable approximation in the high-energy limit \cite{Gyulassy:1999zd}. In addition, the opacity expansion technique developed by Gyulassy, Levai, and Vitev (GLV) is utilized to simplify the analysis of multiple scattering \cite{Gyulassy:2000fs, Gyulassy:2000er}. 
In the present work, we consider a typical scenario in which an energetic quark jet traverses the QCD plasma along a direction perpendicular to the reaction plane. And we seek to explore the influence of the magnetic field on quark dynamics by replacing the vacuum propagator with that in a magnetic field through an appropriate choice of gauge. Since the magnetic field strength is relatively small compared to the jet energy, the weak-field expansion method of propagator is adopted to simplify the calculation, while the higher order terms of the magnetic field are neglected \cite{Chyi:1999fc}. Extending previous analysis, we perform a systematic derivation of the medium-induced photon emission rate for jets of various energies and the corresponding electromagnetic energy loss in the presence of background magnetic fields, and examine how, and to what extent, these quantities are modified as the field strength varies.

The paper is organized as follows. In Sect.~\ref{sect:2}, we introduce the charged scalar propagator under a magnetic field in order to apply it to the calculation below. In Sect.~\ref{sect:3}, we derive medium-induced photon radiation formula and electromagnetic energy loss rate of a quark jet in the GLV formalism. The corresponding numerical analysis and the influence of magnetic field on the above process are given in Sect.~\ref{sect:4}. Finally, we make a brief summary in Sect.~\ref{sect:5}.

\section{\label{sect:2}Charged scalar propagator in a constant magnetic field and weak field approximation}

In the GW model, the spin of the high-energy quark is ignored and the scalar propagator is approximately applied to the calculation of the scattering amplitude \cite{Gyulassy:1999zd}. In this section, we introduce the scalar propagator under a constant background magnetic field, so the effect of the magnetic field on the dynamics of the quark jet is brought in.
A common way to add the magnetic effect into the charged scalar propagator is to use the Schwinger’s proper-time method by considering magnetic field as a background field \cite{Schwinger:1951nm}. For convenience, the direction of a constant magnetic field $B$ is generally set along the $z$-axis and the derivation is calculated in a symmetric gauge: $A_\mu = (0, 0, Bx, 0)$. By choosing that gauge, the Schwinger phase factor can be disregarded \cite{Chyi:1999fc}.
The resummed propagator can be represented in the form of summation over the Landau levels \cite{Ayala:2004dx}
\begin{align}
	i\Delta(p) = 2i \sum_{l=0}^{\infty} \frac{(-1)^l L_l \left(\frac{2p_{\perp}^2}{|qB|}\right)e^{-\frac{p_{\perp}^2}{|qB|}}}
	{p_{\parallel}^2-\left(2l+1\right)|qB|-m^2+i\epsilon},
\end{align}
with $(a\cdot b)_{\parallel}=a^0b^0-a^3b^3$ and $(a\cdot b)_{\perp}=a^1b^1+a^2b^2$.

In the strong field regime, where the magnetic field strength $qB$ significantly exceeds the characteristic momenta, the first term dominates the contribution. This is called the lowest Landau level approximation \cite{Zhang:2024htn}. On the contrary, this expression can be expanded by weak field approximation if we require the hierarchy $qB\ll p^2$ or $qB\ll m^2$ \cite{Chyi:1999fc,Ayala:2004dx}. For a massive hard jet in relativistic heavy-ion collisions, the power counting enables us to use the weak field expansion and rewrite the above equation as \cite{Ayala:2004dx}
\begin{align}
	i\Delta(p) &= 2i\frac{e^{-\frac{p_{\perp}^2}{|qB|}}}{p_{\parallel}^2-m^2} \sum_{l=0}^{\infty}
	\frac{(-1)^l L_l \left(\frac{2p_{\perp}^2}{|qB|}\right)}{1 -\left(2l+1\right)|qB|/(p_{\parallel}^2-m^2)}\nonumber\\
	&=\frac{i}{p_{\parallel}^2-m^2}\sum_{j=0}^{\infty} \bigg(\frac{|qB|}{p_{\parallel}^2-m^2}\bigg)^jS_j,
\end{align}
where
\begin{align}
	S_j &= 2e^{-\frac{p_{\perp}^2}{|qB|}}\sum_{l=0}^{\infty}(-1)^l L_l\left(\frac{2p_{\perp}^2}{|qB|}\right)(2l+1)^j\nonumber\\
    &= i^j \frac{\mathrm{d}^j}{\mathrm{d}x_j}\bigg[ 2e^{-\frac{p_{\perp}^2}{|qB|}}\sum_{l=0}^{\infty}(-1)^l L_l\left(\frac{2p_{\perp}^2}{|qB|}\right)e^{-i(2l+1)x} \bigg] \bigg|_{x=0}\nonumber\\
    &= i^j \frac{\mathrm{d}^j}{\mathrm{d}x_j}\bigg( \frac{e^{-i\frac{p_{\perp}^2}{|qB|}\tan x}}{\cos x} \bigg) \bigg|_{x=0}.
\end{align}

By differentiating with respect to $x$ and substituting $S_j$ back into the expression of propagator, we can simplify and expand the propagator to the quadratic term of $qB$ as \cite{Ayala:2004dx,Ayala:2006sv}
\begin{align}
		i\Delta(p) &= \frac{i}{p_{\parallel}^2-p_{\perp}^2-m^2}\nonumber\\
		&\times \left[1-\frac{(qB)^2}{(p_{\parallel}^2-p_{\perp}^2-m^2)^2}
		-\frac{2(qB)^2p_{\perp}^2}{(p_{\parallel}^2-p_{\perp}^2-m^2)^3} \right].
\end{align}
Rewrite it in the form of four-momentum,
\begin{align}
		i\Delta(p) = \frac{i}{p^2-m^2} \left[1-\frac{(qB)^2}{(p^2-m^2)^2}
		-\frac{2(qB)^2p_{\perp}^2}{(p^2-m^2)^3} \right].
\end{align}

In the following derivation, we will use this form of the propagator throughout. By employing the scalar propagator in a magnetic field, we expect to naturally generalize the GLV energy loss model to the magnetized case.

\section{\label{sect:3}Medium-induced photon radiation in the GLV opacity expansion}
Consider the multiple elastic scatterings of an energetic jet parton with an ensemble of static target partons located at $\boldsymbol{x}_n=(z_n,\bm{x}_{\perp n})$. Assume the scattering centers are far enough apart so that $(z_n-z_{n-1})\gg 1/\mu$, where $\mu$ is the Debye screening mass in hot medium. Then we can model the interactions between the jet and target partons in the QGP by the static color-screened Yukawa potentials \cite{Gyulassy:1993hr,Gyulassy:1999zd}
\begin{align}
	V_n &= V(q_n)e^{iq_n\cdot x_n}\nonumber\\
	&= 2\pi\delta(q^0)v(\bm{q}_n)e^{-i\bm{q}_n\cdot \bm{x}_n}T_{a_n}(j)\otimes T_{a_n}(n),
\end{align}
where $\bm{q}_n$ is the momentum transfer from $n^{\mathrm{th}}$ target and
\begin{align}
v(\bm{q}_n) = \frac{4\pi\alpha_s}{\bm{q}_n^2+\mu^2} = \frac{4\pi\alpha_s}{q_{nz}^2+\mu_n^2},
\end{align}
where $\alpha_s$ the strong coupling constant. While $T_{a_n}(j)$ and $ T_{a_n}(n)$ represent the color matrices for jet and target partons, respectively.

According to Ref.~\cite{Gyulassy:2000er}, we assume that all target partons are in the same $d_T$ dimensional representation with the Casimir operator $C_2(T)$ (${\mathrm{Tr}}T_a(n) = 0$ and ${\mathrm{Tr}}(T_{a}(i)T_{b}(j))=\delta_{ij}\delta_{ab}C_2(T)d_T/d_A$).
In the following, we omit the direct product symbol and shorthand the generators in the $d_R$ dimensional representation corresponding to the jet by $a \equiv t_a$ ($aa = C_R\bm{1}$ and ${\mathrm{Tr}}(aa) = C_Rd_R$).
The elastic cross section between the jet and target partons in the GW model is
\begin{align}\label{eq:3}
	\frac{\mathrm{d}^2\sigma_{\mathrm{el}}}{\mathrm{d}^2 \bm{q}_{\perp}} = \frac{C_RC_2(T)}{d_A}\frac{|v(0,\bm{q}_{\perp})|^2}{(2\pi)^2}.
\end{align}

Our results of photon emission rate are computed up to first order in opacity, and also to the leading order in both $\alpha_e$ and $\alpha_s$. The scattering amplitudes for the first three orders, namely self-quenching, single rescattering and double Born rescattering are denoted as $\mathcal{M}_0$, $\mathcal{M}_1$ and $\mathcal{M}_2$, respectively. Then, the squared total amplitude is given by
\begin{align}
	{|\mathcal{M}|}^2 &= {|\mathcal{M}_0 + \mathcal{M}_1 + \mathcal{M}_2 + \cdots|}^2\nonumber\\
	&= {|\mathcal{M}_0|}^2 + {|\mathcal{M}_1|}^2 + 2{\mathrm{Re}}(\mathcal{M}_1 \mathcal{M}_0^*)\nonumber\\ 
	&\quad +2{\mathrm{Re}}(\mathcal{M}_2 \mathcal{M}_0^*) + \cdots.
\end{align}
Notice that the term about ${\mathrm{Re}}(\mathcal{M}_1 \mathcal{M}_0^*)$ does not contribute to the result due to ${\mathrm{Tr}}T_{a}(j) = 0$.

\subsection{\label{sect:3.1}Spontaneous emission photon of the hard jet}
A wave packet $J(p)$ localized at $x_0=(t_0, \bm{x}_0)$ is used to describe the hard parton jet with momentum $p$ produced in the plasma. The jet will interact with target partons by exchange gluons and subsequently radiates a photon.

The initial and final jet momenta, as well as the momentum and polarization vector of the emitted photon, can then be written in terms of light-cone components as follows:
\begin{align}
	p_i &= \left[p^+,p^-,\bm{0}_{\perp}\right],\nonumber\\
	p_f &= \left[\left(1-x\right)p^+, \frac{\left(\bm{q}_{\perp}-\bm{k}_{\perp}\right)^2 + m^2}{\left(1-x\right)p^+},
	\bm{q}_{\perp}-\bm{k}_{\perp}\right],\nonumber\\
	k &= \left[xp^+, \frac{\bm{k}_{\perp}^{2}}{xp^+}, \bm{k}_{\perp}\right],\nonumber\\
	\epsilon &= \left[0, \frac{2 \bm{\epsilon}_{\perp} \cdot \bm{k}_{\perp}}{xp^+}, \bm{\epsilon}_{\perp}\right],
\end{align}
where $x$ is the momentum fraction of the jet parton that transferred to the photon. For theoretical tractability within our framework, we only consider a kinematic configuration where the quark propagates almost parallel to the magnetic field. During the entire scattering process, the transverse momentum is small compared to the longitudinal momentum, allowing the momentum component perpendicular to the field to be neglected to simplify the calculation.

Our calculation takes $m$ as the quark thermal mass given by the resummation of hard thermal loops, $m_{\mathrm{T}}^2 = g_s^2 C_F T^2/8$ \cite{Braaten:1989mz}, where $C_F=4/3$. For the $u$ quark, $q=2e/3$. In addition, we assume that $J$ varies slowly enough with momentum $p$ as in the Ref.~\cite{Gyulassy:1993hr}.

\begin{figure}[htbp]
	\centering
	\includegraphics[width=1.5in,height=0.49in]{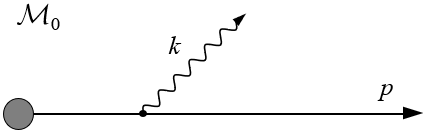}
	\caption{\label{fig:1}Self-quenching photon radiation diagram as the
    direct contributing to the zeroth order in opacity.}
\end{figure}

According to Feynman's rule of GW model, the scattering amplitude of self-quenching photon radiation diagram in Fig.~\ref{fig:1} can be written directly
\begin{align}
\mathcal{M}_0 = -iJ(p+k)e^{i(p+k)\cdot x_0} \mathcal{R}_0,
\end{align}
where we have factored out the part associated with the emitted photon and denoted it as radiation amplitude
\begin{align}
	\mathcal{R}_0 &= -ig_e\epsilon \cdot (2p+k) i\Delta(p+k)\nonumber\\
	&= 2g_e(1-x)\frac{\bm{\epsilon}_{\perp} \cdot \bm{k}_{\perp}}{\bm{k}_{\perp}^{2} + x^2m^2}
    \left[1-\frac{(qB)^2x^2(1-x)^2}{(\bm{k}_{\perp}^{2} + x^2m^2)^2} \right]\nonumber\\
    &= 2g_e(1-x)\bm{\epsilon}_{\perp} \cdot \left[\bm{H} - x^2(1-x)^2\bm{H}^B\right].
\end{align}
Here we define the following quantities for follow the convention of relevant references,
\begin{align}
	\bm{H} &= \frac{\bm{k}_{\perp}}{\bm{k}_{\perp}^2 + x^2m^2},\nonumber\\
    \bm{H}^B &= \frac{(qB)^2\bm{k}_{\perp}}{(\bm{k}_{\perp}^2 + x^2m^2)^3}.
\end{align}
In the zero magnetic field limit, $B=0$, the expression of $\mathcal{R}_0$ naturally reduces to Eq.(9) in Ref.~\cite{Zhang:2010hiv}.

Denote $N_{\gamma}$ as the number of the radiated photon. For the zeroth order in opacity (No interaction between jet and medium parton), the radiation spectrum can be extracted as \cite{Zhang:2010hiv}
\begin{align}
	|\mathcal{M}_0|^2 \frac{\mathrm{d}^3\bm{p}}{(2\pi)^3 2E} \frac{\mathrm{d}^3\bm{k}}{(2\pi)^3 2\omega} = \mathrm{d}^3N_{J} \mathrm{d}^3N_{\gamma}^{(0)},
\end{align}
where the probability distribution for the jet production is given by \cite{Gyulassy:2000er}
\begin{align}
	\mathrm{d}^3N_{J} = d_R|J(p)|^2 \frac{\mathrm{d}^3\bm{p}}{(2\pi)^3 2p^0},
\end{align}
with $d_R = 3$ dimensional representation quarks. And the photon radiation spectrum is
\begin{align}
	\mathrm{d}^3N_{\gamma}^{(0)} = |\mathcal{R}_0|^2 \frac{\mathrm{d}^3\bm{k}}{(2\pi)^3 2\omega},
\end{align}
it can be written as
\begin{align}
	\frac{\mathrm{d}^2 N_{\gamma}^{(0)}}{\mathrm{d}|\bm{k}_{\perp}|^2\mathrm{d}x} = \frac{1}{4(2\pi)^2} \frac{1}{x} |\mathcal{R}_0|^2.
\end{align}

\subsection{\label{sect:3.2}Photon radiation induced by rescattering}
\begin{figure}[htbp]
	\centering
	\includegraphics[width=3.29in,height=0.85in]{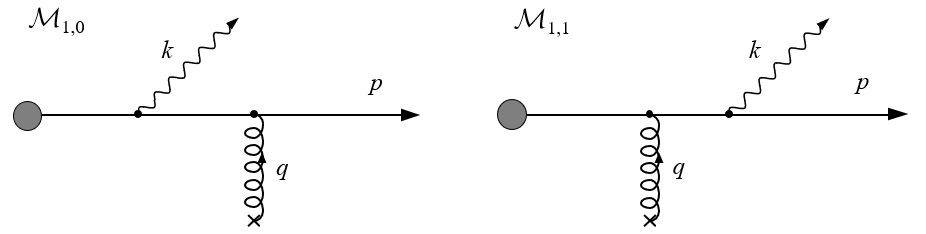}
	\caption{\label{fig:2}Single scattering photon radiation diagrams as the direct contributing to the first order in opacity.}
\end{figure}

In order to compute results of the first order in opacity, we also need to consider the single scattering and Double Born scattering processes. The diagrams describing single scattering photon radiation is shown in Fig.~\ref{fig:2}, with 'cross' stand for the field produced by static partons of medium.

The scattering amplitude corresponding to the first single rescattering diagram is
\begin{widetext}
\begin{align}\label{M10}
	\mathcal{M}_{1,0} &=\int \frac{\mathrm{d}^4 q_1}{(2\pi)^4}iJ(p+k-q_1)e^{i(p+k-q_1)\cdot x_0}
	i\Delta(p+k-q_1) ig_e \epsilon \cdot (2p+k-2q_1) \nonumber\\
	&\quad \times  i\Delta(p-q_1)(-i)(2p-q_1)^0 V(q_1)e^{iq_1\cdot x_1} \nonumber\\
	&\approx J(p+k)e^{i(p+k)\cdot x_0}(-i)g_e 2(E-\omega)\frac{2\bm{\epsilon}_{\perp} \cdot \bm{k}_{\perp}}{x}
	\int \frac{\mathrm{d}^2 \bm{q}_{1\perp}}{(2\pi)^2}e^{-i\bm{q}_{1\perp}\cdot \bm{b}_{1\perp}} \nonumber\\
	&\quad \times \int \frac{\mathrm{d}q_{1z}}{2\pi} v(q_{1z},\bm{q}_{1\perp}) \Delta(p+k-q_1)\Delta(p-q_1)e^{-iq_{1z}(z_1-z_0)} a_1T_{a_1} \nonumber\\
	&=-iJ(p+k)e^{i(p+k)\cdot x_0} \int \frac{\mathrm{d}^2 \bm{q}_{1\perp}}{(2\pi)^2} v(0,\bm{q}_{1\perp})
	e^{-i\bm{q}_{1\perp} \cdot \bm{b}_{1\perp}} a_1T_{a_1}\mathcal{R}_{1,0},
\end{align}
\end{widetext}
where $\bm{b}_{1\perp} = \bm{x}_{1\perp} - \bm{x}_{0\perp}$. We have factored out the radiation part $\mathcal{R}_{1,0}$ from the scattering amplitude here.
In the last line of Eq.~(\ref{M10}), we closed the contour in the lower half of the complex $q_{1z}$ plane and integrated $q_{1z}$.
Here we neglect the exponentially suppressed pole at $-i\mu_1$, under the assumption that $z_1-z_0\gg 1/\mu$. The detailed derivation of $\mathcal{M}_{1,0}$ and calculation about contour integral of $q_{1z}$ are placed in the Appendix~\ref{appe:1}. The expression of the radiation amplitude is given by
\begin{align}
	\mathcal{R}_{1,0} &= -2ig_e(1-x)\frac{\bm{\epsilon}_{\perp} \cdot \bm{k}_{\perp}}{\bm{k}_{\perp}^{2} + x^2m^2}\nonumber\\
	&\quad \times \left[1 + \frac{(qB)^2x^2}{(\bm{k}_{\perp}^{2} + x^2m^2)^2} 
	+ \frac{(qB)^2x^2(1-x)^2}{(\bm{k}_{\perp}^{2} + x^2m^2)^2}\right]\nonumber\\
    &\quad \times \big[1-e^{i\omega_0(z_1-z_0)}\big],
\end{align}
where we define
\begin{align}
	\omega_0 = \frac{\bm{k}_{\perp}^2 + x^2m^2}{2\omega(1-x)}.
\end{align}

By doing a similar calculation, we can also obtain the radiation amplitude of second single scattering diagram,
\begin{align}
	\mathcal{R}_{1,1} &= -2ig_e(1-x)\frac{\bm{\epsilon}_{\perp} \cdot (\bm{k}_{\perp}-x\bm{q}_{1\perp})}{(\bm{k}_{\perp}-x\bm{q}_{1\perp})^2 + x^2m^2}\nonumber\\
    &\quad \times \left\{1-\frac{(qB)^2x^2(1-x)^2}{[(\bm{k}_{\perp}-x\bm{q}_{1\perp})^2 + x^2m^2]^2} \right\}\nonumber\\
    &\quad \times e^{i\omega_0(z_1-z_0)}.
\end{align}

Add the two together to obtain the single scattering radiation amplitude
\begin{align}
	\mathcal{R}_1 &= \mathcal{R}_{1,0} + \mathcal{R}_{1,1}\nonumber\\
	&= -2ig_e(1-x)\bm{\epsilon}_{\perp}\cdot \big\{ \big[\bm{H} + x^2\bm{H}^B + x^2(1-x)^2\bm{H}^B\big]\nonumber\\
    &\quad \times \big[1-e^{i\omega_0(z_1-z_0)}\big]\nonumber\\
    &\quad + \big[\bm{C}_1 - x^2(1-x)^2\bm{C}_1^B\big] e^{i\omega_0(z_1-z_0)}\big\},
\end{align}
where
\begin{align}
	\bm{C}_1 &= \frac{\bm{k}_{\perp}-x\bm{q}_{1\perp}}{(\bm{k}_{\perp}-x\bm{q}_{1\perp})^2 + x^2m^2},\nonumber\\
	\bm{C}_1^B &= \frac{(qB)^2(\bm{k}_{\perp}-x\bm{q}_{1\perp})}{[(\bm{k}_{\perp}-x\bm{q}_{1\perp})^2 + x^2m^2]^3}.
\end{align}

\begin{figure}[htbp]
	\centering
	\includegraphics[width=3.29in,height=0.85in]{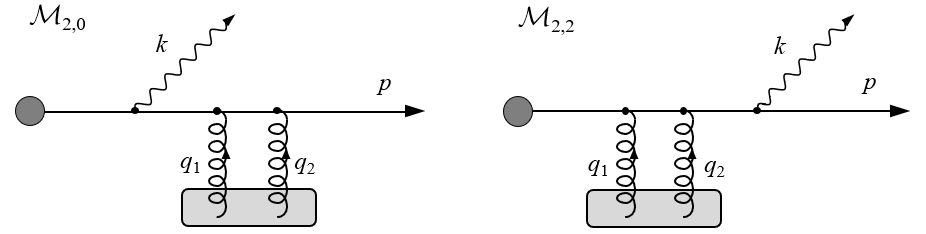}
	\caption{\label{fig:3}Double Born scattering photon radiation diagrams as the contact-limit contributing to the first order in opacity.}
\end{figure}

The diagrams describing the double Born scattering photon radiation is illustrated in Fig.~\ref{fig:3}, where the 'shaded boxes' indicate the contact limit of two scattering centers. In the framework of time-ordered perturbation theory, the contribution from diagram representing the photon radiated in between the two scattering centers can be proved to vanish \cite{Gyulassy:2000er}, hence the diagram $\mathcal{M}_{2,1}$ is not drawn here.

The scattering amplitude corresponding to the first double rescattering diagram is
\begin{widetext}
\begin{align}\label{M20}
	\mathcal{M}_{2,0} &=\int \frac{\mathrm{d}^4 q_1}{(2\pi)^4}\int \frac{\mathrm{d}^4 q_2}{(2\pi)^4}iJ(p+k-q_1-q_2)
	e^{i(p+k-q_1-q_2)\cdot x_0} i\Delta(p+k-q_1-q_2) ig_e\epsilon \cdot \left(2p+k-2q_1-2q_2\right)\nonumber\\
	&\quad \times i\Delta(p-q_1-q_2) (-i)(2p-q_1-2q_2)^0 V(q_1)e^{iq_1\cdot x_1}
		i\Delta(p-q_2)(-i)(2p-q_2)^0 V(q_2)e^{iq_2\cdot x_2}\nonumber\\
	&\approx J(p+k)e^{i(p+k)\cdot x_0}(-i)g_e 4(E-\omega)^2\frac{2\bm{\epsilon}_{\perp} \cdot \bm{k}_{\perp}}{x}\nonumber\\
	&\quad \times \int \frac{\mathrm{d}^2 \bm{q}_{1\perp}}{(2\pi)^2}e^{-i\bm{q}_{1\perp}\cdot \bm{b}_{1\perp}}
				\int \frac{\mathrm{d}^2 \bm{q}_{2\perp}}{(2\pi)^2}e^{-i\bm{q}_{2\perp}\cdot \bm{b}_{2\perp}}
				\int \frac{\mathrm{d}q_{2z}}{2\pi} v(q_{2z},\bm{q}_{2\perp})\Delta(p-q_2) e^{-iq_{2z}(z_2-z_1)}\nonumber\\
    &\quad \times \int \frac{\mathrm{d}q_{1z}}{2\pi} v(q_{1z},\bm{q}_{1\perp}) \Delta(p+k-q_1-q_2) \Delta(p-q_1-q_2) e^{-i(q_{1z}+q_{2z})(z_1-z_0)} a_2a_1T_{a_2}T_{a_1}.
\end{align}
\end{widetext}
In the last line of Eq.~(\ref{M20}), we will integrate over $q_{1z}$ by closing the contour in lower half-plane and neglect the pole at $-i\mu_1$ again, and set $x_1=x_2=x_j$ here for the contact limit.
Then close the contour in the lower half of the complex $q_{2z}$ plane and integrate over $q_{2z}$. Note that since there is no exponentially suppressed factor, the poles at $q_{2z}=-i\mu_1$ and $-i\mu_2$ also contribute to the integral. The detailed derivation of $\mathcal{M}_{2,0}$ and calculations about contour integral over $q_{1z}$ and $q_{2z}$ are placed in the Appendix~\ref{appe:1}. Finally, the expression is given by
\begin{align}
	\mathcal{M}_{2,0} &=-iJ(p+k)e^{i(p+k)\cdot x_0}\nonumber\\
	&\quad \times \int \frac{\mathrm{d}^2 \bm{q}_{1\perp}}{(2\pi)^2} v(0,\bm{q}_{1\perp})
	\int \frac{\mathrm{d}^2 \bm{q}_{2\perp}}{(2\pi)^2} v(0,\bm{q}_{2\perp})\nonumber\\
	&\quad \times e^{-i(\bm{q}_{1\perp}+\bm{q}_{2\perp})\cdot \bm{b}_{j\perp}}a_2a_1T_{a_2}T_{a_1}\mathcal{R}_{2,0},
\end{align}
where
\begin{align}
    \mathcal{R}_{2,0} &= -g_e(1-x)\bm{\epsilon}_{\perp}\nonumber\\
    &\quad \cdot \left[\bm{H} + x^2\bm{H}^B + x^2(1-x)^2\bm{H}^B\right]\nonumber\\
    &\quad \times \big[1- e^{i\omega_0(z_j-z_0)}\big]
\end{align}
is the corresponding radiation amplitude.

By doing a similar calculation, we can obtain the radiation amplitude of the second double Born scattering diagram
\begin{align}
    \mathcal{R}_{2,2} &= -g_e(1-x)\bm{\epsilon}_{\perp}\cdot
    \left[\bm{C}_2 - x^2(1-x)^2\bm{C}_2^B\right]\nonumber\\
    &\quad \times e^{i\omega_0(z_j-z_0)},
\end{align}
where
\begin{align}
    \bm{C}_2 &= \frac{\bm{k}_{\perp}-x\bm{q}_{1\perp}-x\bm{q}_{2\perp}}{(\bm{k}_{\perp} 
    - x\bm{q}_{1\perp} - x\bm{q}_{2\perp})^2 + x^2m^2},\nonumber\\
    \bm{C}_2^B &= \frac{(qB)^2(\bm{k}_{\perp}-x\bm{q}_{1\perp}-x\bm{q}_{2\perp})}{[(\bm{k}_{\perp} 
    - x\bm{q}_{1\perp} - x\bm{q}_{2\perp})^2 + x^2m^2]^3}.
\end{align}

Add the two together to obtain the double Born scattering radiation amplitude
\begin{align}
	\mathcal{R}_2 &= \mathcal{R}_{2,0} + \mathcal{R}_{2,2}\nonumber\\
	&= -g_e(1-x)\bm{\epsilon}_{\perp}\cdot \big\{ \big[\bm{H} + x^2\bm{H}^B + x^2(1-x)^2\bm{H}^B\big]\nonumber\\
    &\quad \times \big[1- e^{i\omega_0(z_j-z_0)}\big]\nonumber\\
    &\quad + \big[\bm{C}_2 - x^2(1-x)^2\bm{C}_2^B\big] e^{i\omega_0(z_j-z_0)} \big\}.
\end{align}

After squaring the rescattering amplitude, we also have to perform the ensemble averaging for each time scattering process. If the relative transverse coordinate, or impact parameter $\bm{b}_{j\perp} = \bm{x}_{j\perp} - \bm{x}_{0\perp}$, varies over a large transverse area $A_{\perp}$ as compared to the interaction area $1/\mu^2$, the ensemble average over the scattering center location can be expressed as the combination of the impact parameter average and the longitudinal locations average \cite{Gyulassy:2000fs}, as follows
\begin{align}
	\left \langle \cdots \right \rangle = \frac{1}{A_{\perp}} \int \mathrm{d}^2 \bm{b}_{j\perp} \mathrm{d}z_0\mathrm{d}z_j \rho(z_0,z_j)\cdots,
\end{align}
where performing the average over the impact parameter on the phase factor gives $(2\pi)^2 \delta^2(\bm{q}_{\perp}-\bm{q}'_{\perp})$.

Define $L$ as the target size. Longitudinally, the distribution of the initial location of jet and the location of scattering center is defined by \cite{Zhang:2010hiv}
\begin{align}
	\rho(z_0,z) &= \frac{\theta(L-z)}{L/2}{\mathrm{exp}} \bigg( -\frac{L-z}{L/2} \bigg)\nonumber\\
	&\quad \times \frac{\theta(z-z_0)}{L/2}{\mathrm{exp}} \bigg( -\frac{z-z_0}{L/2} \bigg).
\end{align}

For the part about single rescattering, carry out the ensemble and color averages and use Eq.~(\ref{eq:3}) give
\begin{align}
	{\mathrm{Tr}}\left \langle |\mathcal{M}_1|^2 \right \rangle &= d_Rd_T|J(p)|^2\int\mathrm{d}z_0\int\mathrm{d}z \rho(z_0,z)\nonumber\\
	&\quad \times \frac{N}{A_{\perp}} \int \mathrm{d}^2 \bm{q}_{\perp}\frac{C_RC_2(T)}{d_A}
		\frac{|v(0,\bm{q}_{\perp})|^2}{(2\pi)^2}|\mathcal{R}_1|^2\nonumber\\
	&= d_Rd_T|J(p)|^2\int\mathrm{d}z_0\int\mathrm{d}z \rho(z_0,z)\nonumber\\
	&\quad \times \frac{N}{A_{\perp}} \int \mathrm{d}^2 \bm{q}_{\perp} \frac{\mathrm{d}^2\sigma_{\mathrm{el}}}{\mathrm{d}^2 \bm{q}_{\perp}}|\mathcal{R}_1|^2,
\end{align}
here ${\mathrm{Tr}}(T_{a}T_{b}) = \delta_{ab}C_2(T)d_T/d_A$ and ${\mathrm{Tr}}(aa) = C_Rd_R$ have been used. The factor $N$ comes from summing over all the scattering centers.
Making the following rewriting,
\begin{align}
	\frac{N}{A_{\perp}}\int \mathrm{d}^2 \bm{q}_{\perp}\frac{\mathrm{d}^2\sigma_{\mathrm{el}}}{\mathrm{d}^2 \bm{q}_{\perp}}
	=\frac{N\sigma_{\mathrm{el}}}{A_{\perp}}\int \mathrm{d}^2 \bm{q}_{\perp}\frac{1}{\sigma_{\mathrm{el}}}\frac{\mathrm{d}^2\sigma_{\mathrm{el}}}{\mathrm{d}^2 \bm{q}_{\perp}},
\end{align}
we can find 
\begin{align}
	\frac{N\sigma_{\mathrm{el}}}{A_{\perp}} = \frac{L}{\lambda} \equiv \bar{n},
\end{align}
where $\bar{n}$ is called the opacity, which represents the mean number of rescatterings. We can define $|\bar{v}(0,\bm{q}_{\perp})|^2$, the normalized distribution of momentum transfers from the scattering centers, as
\begin{align}
\frac{1}{\sigma_{\mathrm{el}}}\frac{\mathrm{d}^2\sigma_{\mathrm{el}}}{\mathrm{d}^2 \bm{q}_{\perp}}
\equiv |\bar{v}(0,\bm{q}_{\perp})|^2 = \frac{1}{\pi} \frac{\mu^2}{(\bm{q}^2+\mu^2)^2},
\end{align}
The normalization condition for $|\bar{v}(0,\bm{q}_{\perp})|^2$ is enforced by setting $\bm{q}^{2}_{\perp {\max}} = 3E\mu$ in our numerical calculation. So we have
\begin{align}
		{\mathrm{Tr}}\left \langle |\mathcal{M}_1|^2 \right \rangle
		&= d_Rd_T|J(p)|^2\frac{L}{\lambda} \int \mathrm{d}^2 \bm{q}_{\perp} |\bar{v}(0,\bm{q}_{\perp})|^2\nonumber\\
		&\quad \times \int \mathrm{d}z_0 \int \mathrm{d}z \rho(z_0,z)|\mathcal{R}_1|^2.
\end{align}
By a similar analysis, we obtain
\begin{align}
		{\mathrm{Tr}}\left \langle 2{\mathrm{Re}}(\mathcal{M}_2 \mathcal{M}_0^*) \right \rangle
		&= d_Rd_T|J(p)|^2 \frac{L}{\lambda} \int \mathrm{d}^2 \bm{q}_{\perp} |\bar{v}(0,\bm{q}_{\perp})|^2\nonumber\\
		&\quad \times \int \mathrm{d}z_0 \int \mathrm{d}z \rho(z_0,z)2{\mathrm{Re}}(\mathcal{R}_2 \mathcal{R}_0^*).
\end{align}

Notice after taking the impact parameter averaging, $\bm{C}_2$ and $\bm{C}_2^B$ in $\mathcal{M}_2$ will reduce to $\bm{H}$ and $\bm{H}^B$ respectively because of the two-dimensional integration over $\bm{q}_{\perp}$ with the delta function $\delta^2(\bm{q}_{1\perp}+\bm{q}_{2\perp})$. The detailed derivation is placed in the Appendix~\ref{appe:2}. So far, we still need to address
\begin{widetext}
\begin{align}\label{TrM}
	{\mathrm{Tr}}\left \langle |\mathcal{M}_1|^2 + 2{\mathrm{Re}}(\mathcal{M}_2 \mathcal{M}_0^*) \right \rangle
	&= d_Rd_T|J(p)|^2\frac{L}{\lambda}\int \frac{\mu^2\mathrm{d}^2\bm{q}_{\perp}}{\pi(\bm{q}_{\perp}^2 + \mu^2)^2}
		\int \mathrm{d}z_0 \int \mathrm{d}z \rho(z_0,z) \big[|\mathcal{R}_1|^2 + 2{\mathrm{Re}}(\mathcal{R}_2 \mathcal{R}_0^*)\big]\nonumber\\
	&= 4g_e^2 d_Rd_T|J(p)|^2(1-x)^2\frac{L}{\lambda}\int \mathrm{d}^2 \bm{q}_{\perp} |\bar{v}(0,\bm{q}_{\perp})|^2
		\int \mathrm{d}z_0 \int \mathrm{d}z \rho(z_0,z)\nonumber\\
	&\quad \times \big\{(\bm{H} - \bm{C}_1)^2 + x^2(3 + 4(1-x)^2)\bm{H}\cdot\bm{H}^B\nonumber\\
	&\quad + 2x^2(1-x)^2(\bm{H}\cdot\bm{C}_1^B - \bm{C}_1\cdot\bm{C}_1^B) - 2x^2(1 + (1-x)^2)\bm{H}^B\cdot\bm{C}_1\nonumber\\
	&\quad -\cos [\omega_0(z-z_0)] \big[2(\bm{H}^2-\bm{H} \cdot \bm{C}_1) + x^2(3 + 2(1-x)^2)\bm{H}\cdot\bm{H}^B\nonumber\\
	&\quad + 2x^2(1-x)^2\bm{H}\cdot\bm{C}_1^B - 2x^2(1 + (1-x)^2)\bm{H}^B\cdot\bm{C}_1 \big] \big\}.
\end{align}
\end{widetext}
In the above expression, the cosine function reveals a destructive interference known as the Abelian Landau-Pomeranchuk-Migdal (LPM) effect \cite{Wang:1994fx}. Note that in the zero magnetic field limit,  this equation reduces to Eq.(34) in Ref.~\cite{Zhang:2010hiv} again.

According to the Ref.~\cite{Djordjevic:2003zk}, the first order in opacity photon emission rate can be computed from formula
\begin{align}
	\mathrm{d}^3N_{J} \mathrm{d}^3N_{\gamma}^{(1)} &= \frac{1}{d_T} {\mathrm{Tr}}\left \langle|\mathcal{M}_1|^2 
	+ 2{\mathrm{Re}}(\mathcal{M}_2 \mathcal{M}_0^*)\right \rangle \nonumber\\
	&\quad \times \frac{\mathrm{d}^3\bm{p}}{(2\pi)^3 2E} \frac{\mathrm{d}^3\bm{k}}{(2\pi)^3 2\omega},
\end{align}
with $d_T=8$ is the dimension of the target color representation for a pure gluon plasma. Then the photon radiation spectrum now is
\begin{align}
	\frac{\mathrm{d}^2N_{\gamma}^{(1)}}{\mathrm{d}|\bm{k}_{\perp}|^2\mathrm{d}x} &= \frac{1}{4(2\pi)^2}\frac{1}{x}
    \frac{L}{\lambda} \int \mathrm{d}^2 \bm{q}_{\perp} |\bar{v}(0,\bm{q}_{\perp})|^2\nonumber\\
	&\quad \times \int \mathrm{d}z_0 \int \mathrm{d}z \rho(z_0,z)\nonumber\\
	&\quad \times \big[|\mathcal{R}_1|^2 + 2{\mathrm{Re}}(\mathcal{R}_2 \mathcal{R}_0^*)\big].
\end{align}
The integration of LPM term over longitudinal location distribution gives
\begin{align}
	I(\omega_0,L) &\equiv \int \mathrm{d}z_0 \int \mathrm{d}z\rho(z_0,z)\cos [\omega_0(z-z_0)]\nonumber\\
	&=\frac{4}{4+\omega_0^2 L^2}.
\end{align}

By defining several dimensionless quantities
\begin{align}
	u = \frac{|{\bm{q}}_\perp|^2}{\mu^2},\quad y = \frac{|{\bm{k}}_\perp|^2}{\mu^2},\quad w = \frac{m^2}{\mu^2},\quad v = \frac{qB}{\mu^2},
\end{align}
the differential expressions for the photon radiation yield at zeroth order and first order in opacity can be simplified to the following forms, respectively:
\begin{align}
	\frac{\mathrm{d}^2N_{\gamma}^{(0)}}{\mathrm{d}x\mathrm{d}y} &= \frac{4\alpha_e}{9\pi} \frac{(1-x)^2}{x}\bigg[ \frac{y}{(y+x^2w)^2} \nonumber\\
    &\quad - 2x^2(1-x)^2 \frac{v^2y}{(y+x^2w)^4} \bigg],
\end{align}
\begin{widetext}
\begin{align}\label{eq.dN_1}
	\frac{\mathrm{d}^2N_{\gamma}^{(1)}}{\mathrm{d}x\mathrm{d}y} &= \frac{4\alpha_e}{9\pi} \frac{(1-x)^2}{x} \frac{L}{\lambda}
            \int \frac{\mathrm{d}\theta}{2\pi} \int \mathrm{d}u\frac{\mu^2}{(1+u)^2}
            \bigg\{ \frac{y}{(y+x^2w)^2} - 2\frac{y-x\sqrt{yu}\cos\theta}{(y+x^2w)(y+x^2u-2x\sqrt{yu}\cos\theta+x^2w)}\nonumber\\
    &\quad + \frac{y+x^2u-2x\sqrt{yu}\cos\theta}{(y+x^2u-2x\sqrt{yu}\cos\theta+x^2w)^2} + x^2(3+4(1-x)^2) \frac{v^2y}{(y+x^2w)^4}\nonumber\\
    &\quad + 2x^2(1-x)^2 \frac{v^2(y-x\sqrt{yu}\cos\theta)}{(y+x^2w)(y+x^2u-2x\sqrt{yu}\cos\theta+x^2w)^3}
            - 2x^2(1-x)^2 \frac{v^2(y+x^2u-2x\sqrt{yu}\cos\theta)}{(y+x^2u-2x\sqrt{yu}\cos\theta+x^2w)^4}\nonumber\\
    &\quad - 2x^2(1+(1-x)^2) \frac{v^2(y-x\sqrt{yu}\cos\theta)}{(y+x^2w)^3(y+x^2u-2x\sqrt{yu}\cos\theta+x^2w)}\nonumber\\
    &\quad - I(\omega_0,L) \bigg[2\frac{y}{(y+x^2w)^2} - 2\frac{y-x\sqrt{yu}\cos\theta}{(y+x^2w)(y+x^2u-2x\sqrt{yu}\cos\theta+x^2w)}
            + x^2(3+2(1-x)^2) \frac{v^2y}{(y+x^2w)^4}\nonumber\\
    &\quad + 2x^2(1-x)^2 \frac{v^2(y-x\sqrt{yu}\cos\theta)}{(y+x^2w)(y+x^2u-2x\sqrt{yu}\cos\theta+x^2w)^3}\nonumber\\
    &\quad - 2x^2(1+(1-x)^2) \frac{v^2(y-x\sqrt{yu}\cos\theta)}{(y+x^2w)^3(y+x^2u-2x\sqrt{yu}\cos\theta+x^2w)} \bigg] \bigg\}.
\end{align}
\end{widetext}
The angle $\theta$ in the second equation is between ${\bm{p}}_\perp$ and ${\bm{k}}_\perp$ and will be integrated out in the final photon radiation spectrum.

For the quark massless limit, when $|{\bm{k}}_\perp| \sim 0$ or $|{\bm{k}}_\perp| \sim x|{\bm{q}}_\perp|$, the above function will have two kinds of collinear divergence, corresponding to the momentum of emitted photon collinear to initial and final jet momentum respectively. Our work takes into account the quark thermal mass that acts as a regulator, so these divergences can be avoided.

\section{\label{sect:4}Jet electromagnetic radiation and energy loss in a magnetic field, numerical result}

\begin{figure}[htbp]
	\centering
	\includegraphics[width=2.9in,height=2.3in]{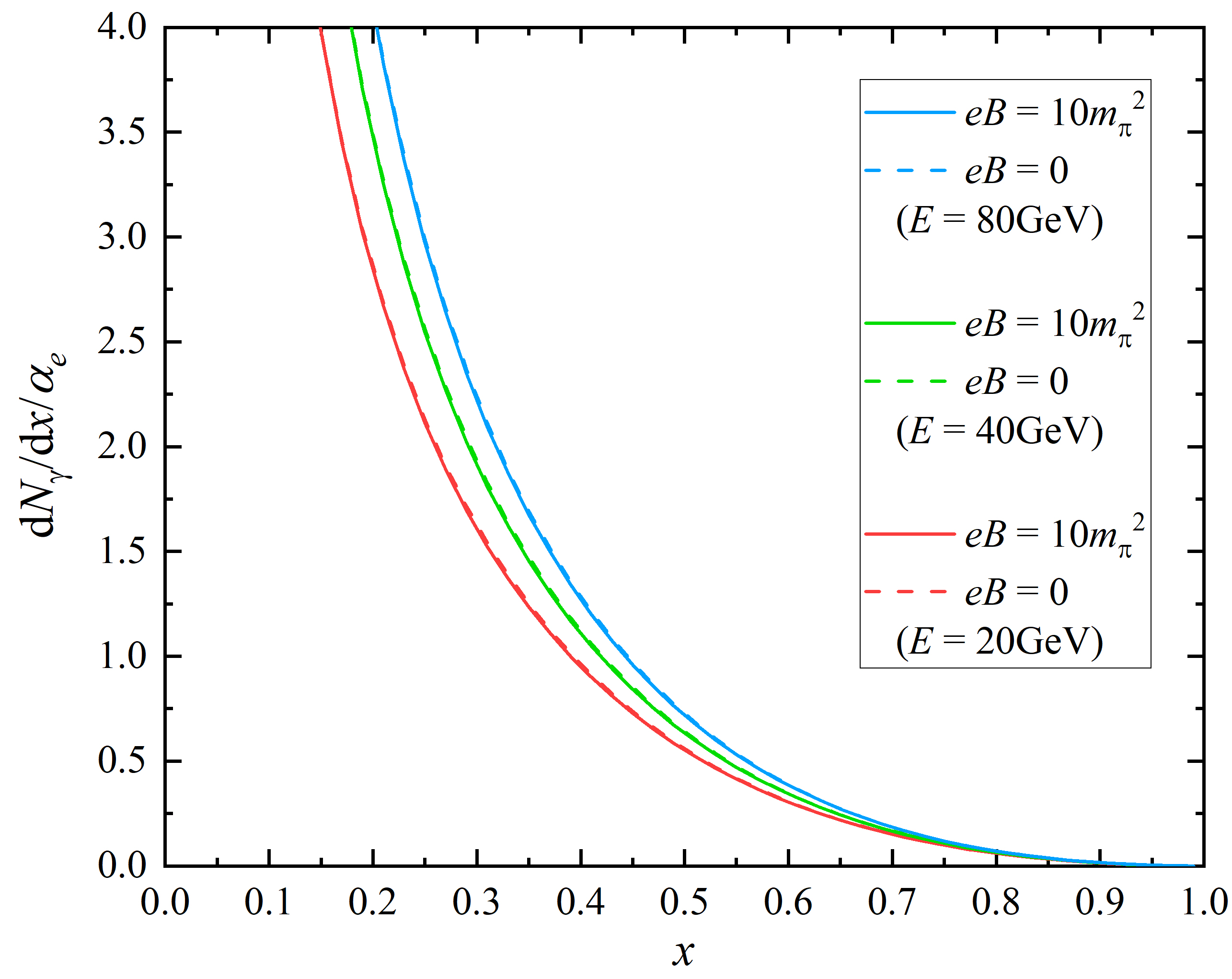}
	\caption{\label{fig:4}Comparison of the photon yields as a function of $x$ for different jet initial energies $E$, with and without magnetic field.}
\end{figure}
\begin{figure}[htbp]
	\centering
	\includegraphics[width=2.9in,height=2.3in]{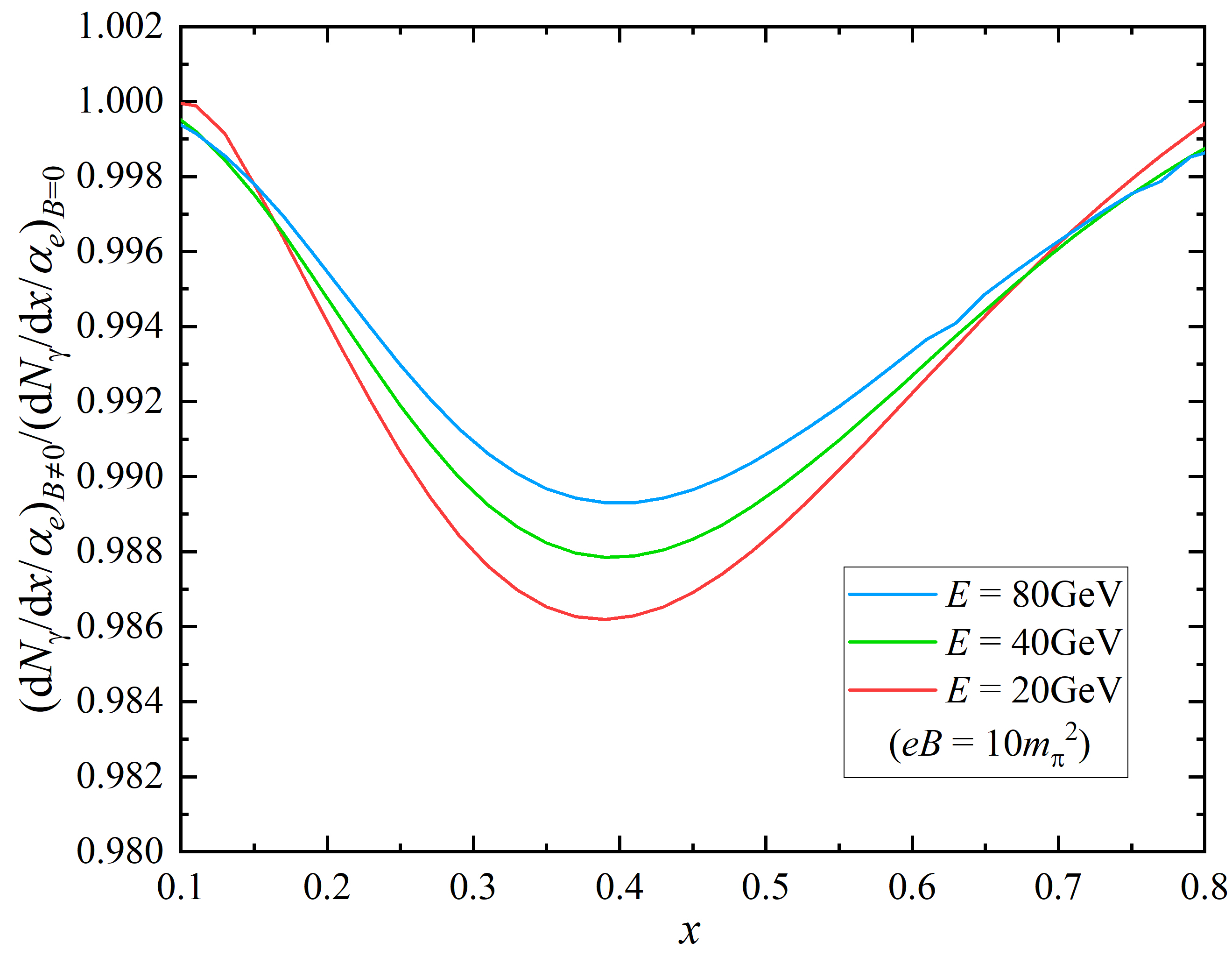}
	\caption{\label{fig:5}Ratios of the photon yields as a function of $x$ with to without magnetic field for different jet initial energies $E$.}
\end{figure}
\begin{figure}[htbp]
	\centering
	\includegraphics[width=2.9in,height=2.3in]{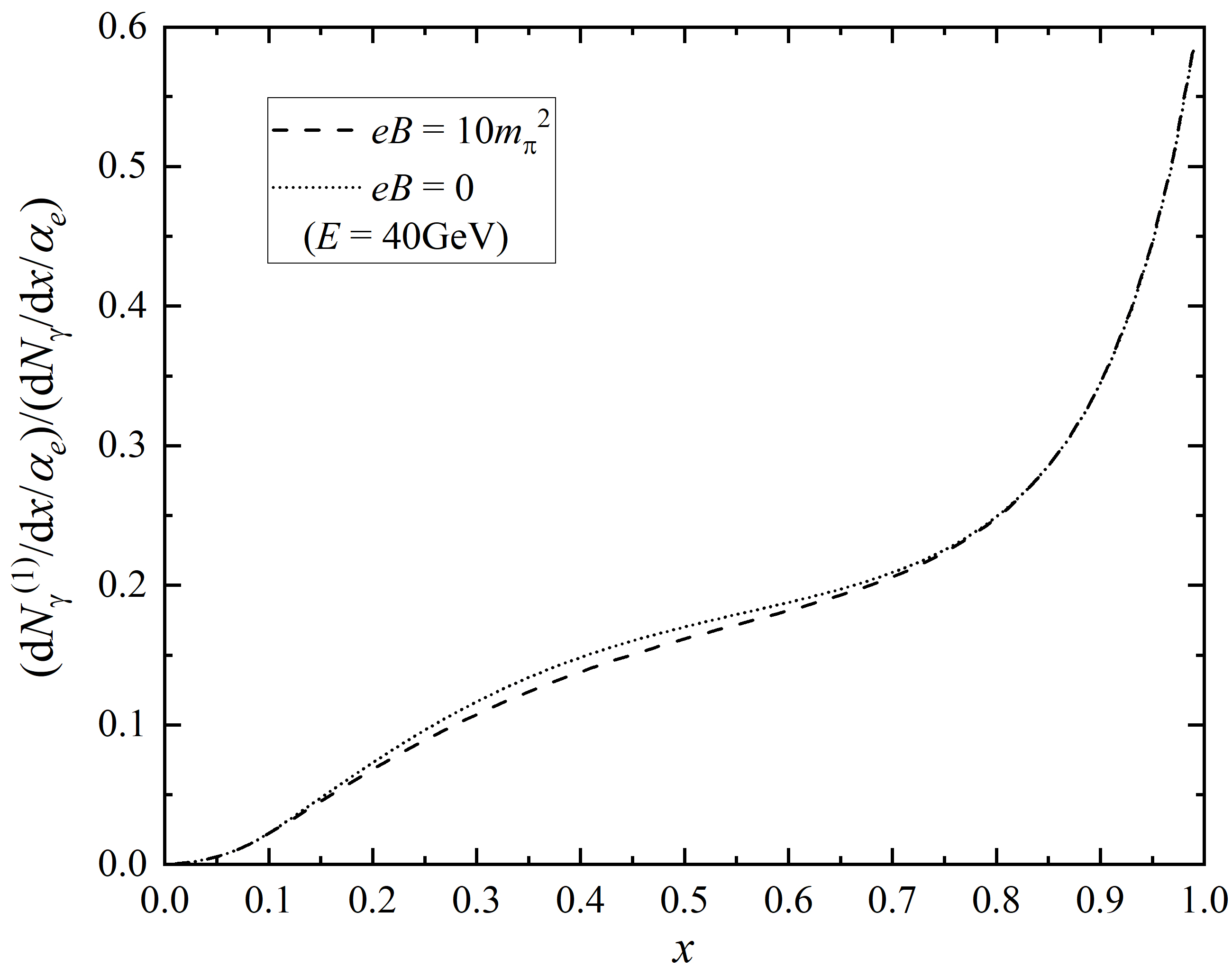}
	\caption{\label{fig:6}Ratios of the first order in opacity to the total photon yield, with and without magnetic field.}
\end{figure}
Denote $N_{\gamma}=N_{\gamma}^{(0)}+N_{\gamma}^{(1)}$ as the total number of radiation photon computed to first order in opacity. To determine the photon production rate as a function of the radiated photon energy, it is necessary to integrate over the transverse momentum squared, ${\bm{k}}_\perp^2$. We take the kinematic limits to be $\mu^2 \le {\bm{k}}_\perp^2 \le 4E^2x(1-x)$ \cite{Wang:2001cs}.

By multiplying photon production rates by $x$ and integrating over ${\bm{k}}_\perp^2$ and $x$ for different value of jet energy, we can obtain the fractional (electromagnetic) energy loss of the jet,
\begin{align}
	\frac{\Delta E}{E} = \int x\mathrm{d}x \int \mathrm{d}\bm{k}_{\perp}^2\frac{\mathrm{d}^2N_{\gamma}}{\mathrm{d}\bm{k}_{\perp}^2 \mathrm{d}x}.
\end{align}

Following Ref.~\cite{Gyulassy:2000er}, we set the screening mass to $\mu^2 = 4\pi \alpha_{s} T^2$ with $\alpha_s=0.3$ in our numerical calculations. To focus on the effects of the magnetic field, we fix the medium length at $L=6fm$ and the mean free path at $\lambda=1fm$, corresponding to opacity $\bar{n}=6$.

Fig.~\ref{fig:4} shows the comparison of photon radiation yields as a function of photon energy fraction $x$ for different jet initial energies $E$ with and without background magnetic field. It can be seen from the figure that the higher the jet energy, the greater the number of emitted photons. This trend is expected, as a higher energy results in more intense interactions with the hot medium during a jet propagation through the plasma. It can be also found that the total photon yield decreases with increasing photon energy $xE$; in other words, most of the photons radiated from the jet due to medium-induced processes are soft. These phenomena are consistent with the conclusions from Ref.~\cite{Zhang:2010hiv}. We also found that the overall photon yield slightly decreases as the jet moves in the direction of the magnetic field.

Fig.~\ref{fig:5} further presents the ratios of photon yield of in the presence to absence of magnetic field for different jet initial energies. Although the magnetic field effect is not very significant, we can still observe that the reduction of photon yield by the background magnetic field is more effective for photons carrying medium and low energy. Moreover, as the jet becomes harder, the process of medium-induced photon radiation becomes less susceptible to the magnetic field.

In order to see the effect of magnetic field on photon bremsstrahlung more carefully, we plot the ratio of the first order in opacity to the total photon production rates in Fig.~\ref{fig:6}. We observe that the contributions from high order in opacity terms remain relatively small in the presence of a background magnetic field. This suggests that our approach for introducing the magnetic field does not disrupt the structure of opacity expansion.
\begin{figure}[htbp]
	\centering
	\includegraphics[width=3.0in,height=2.34in]{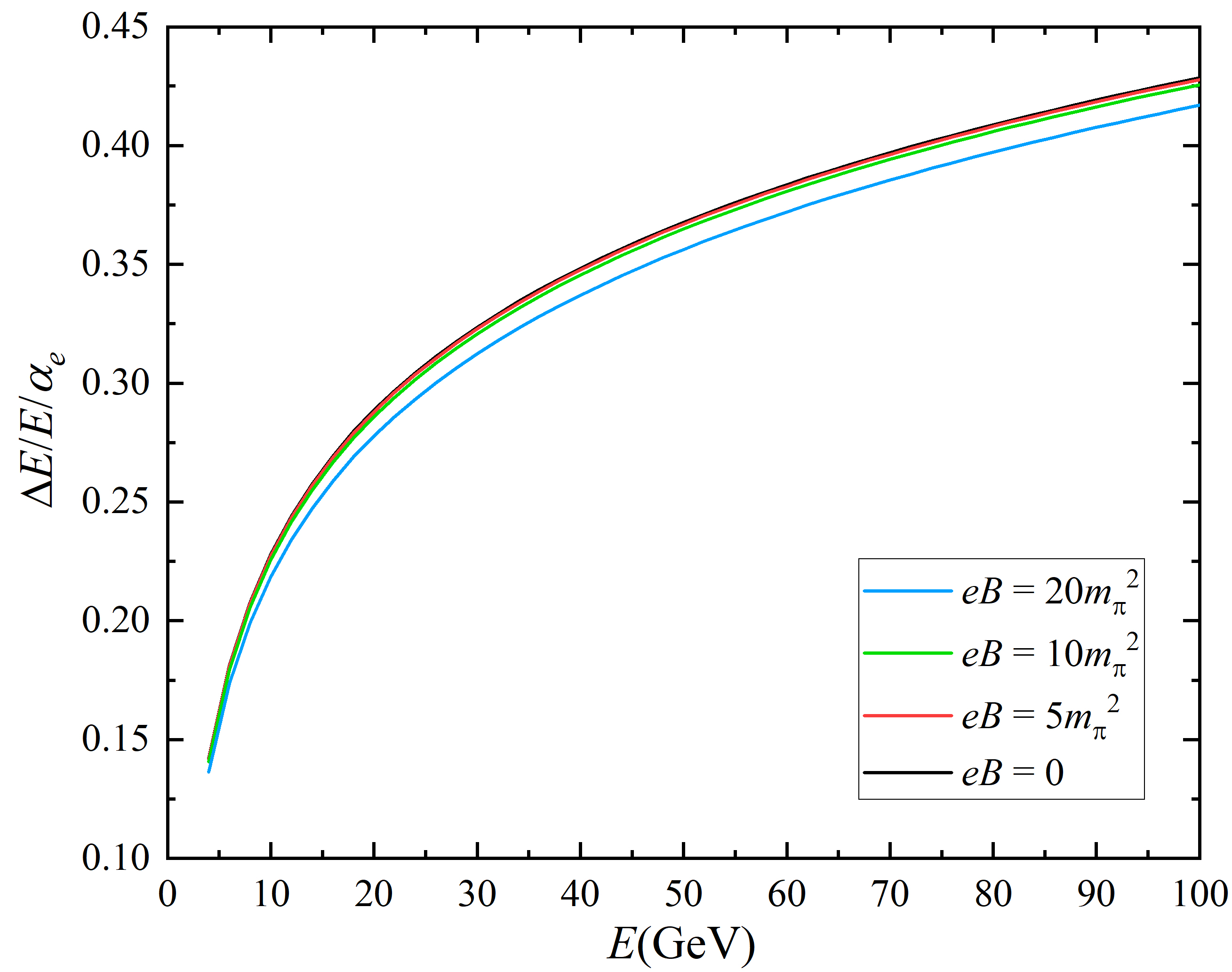}
	\caption{\label{fig:7}Comparison of the jet fractional electromagnetic energy loss as a function of jet initial energy $E$, without and with different magnetic fields.}
\end{figure}
\begin{figure}[htbp]
	\centering
	\includegraphics[width=3.0in,height=2.34in]{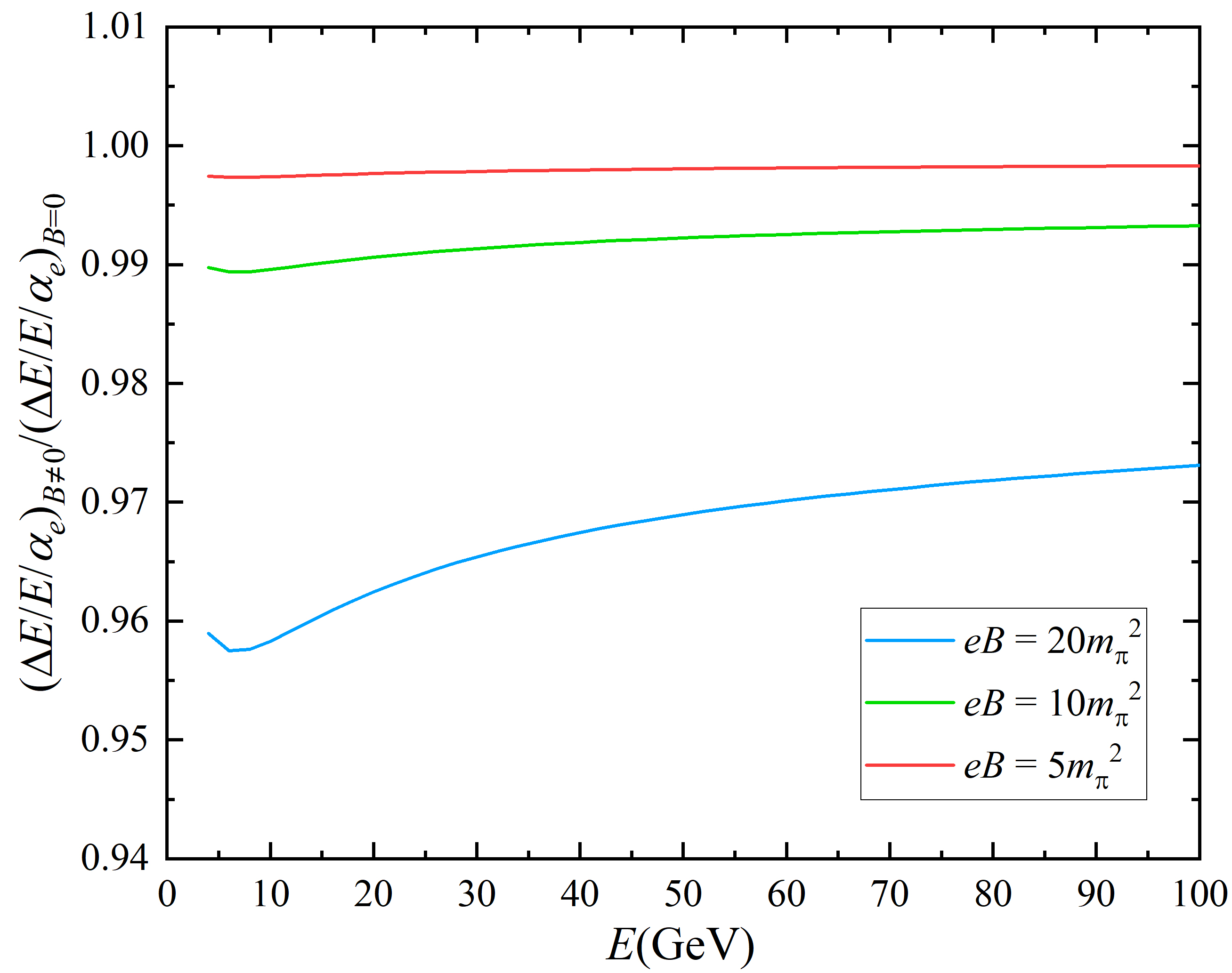}
	\caption{\label{fig:8}Ratios of the jet fractional electromagnetic energy loss as a function of jet initial energy $E$ with different magnetic fields to without magnetic field.}
\end{figure}
\begin{figure}[htbp]
	\centering
	\includegraphics[width=3.0in,height=2.34in]{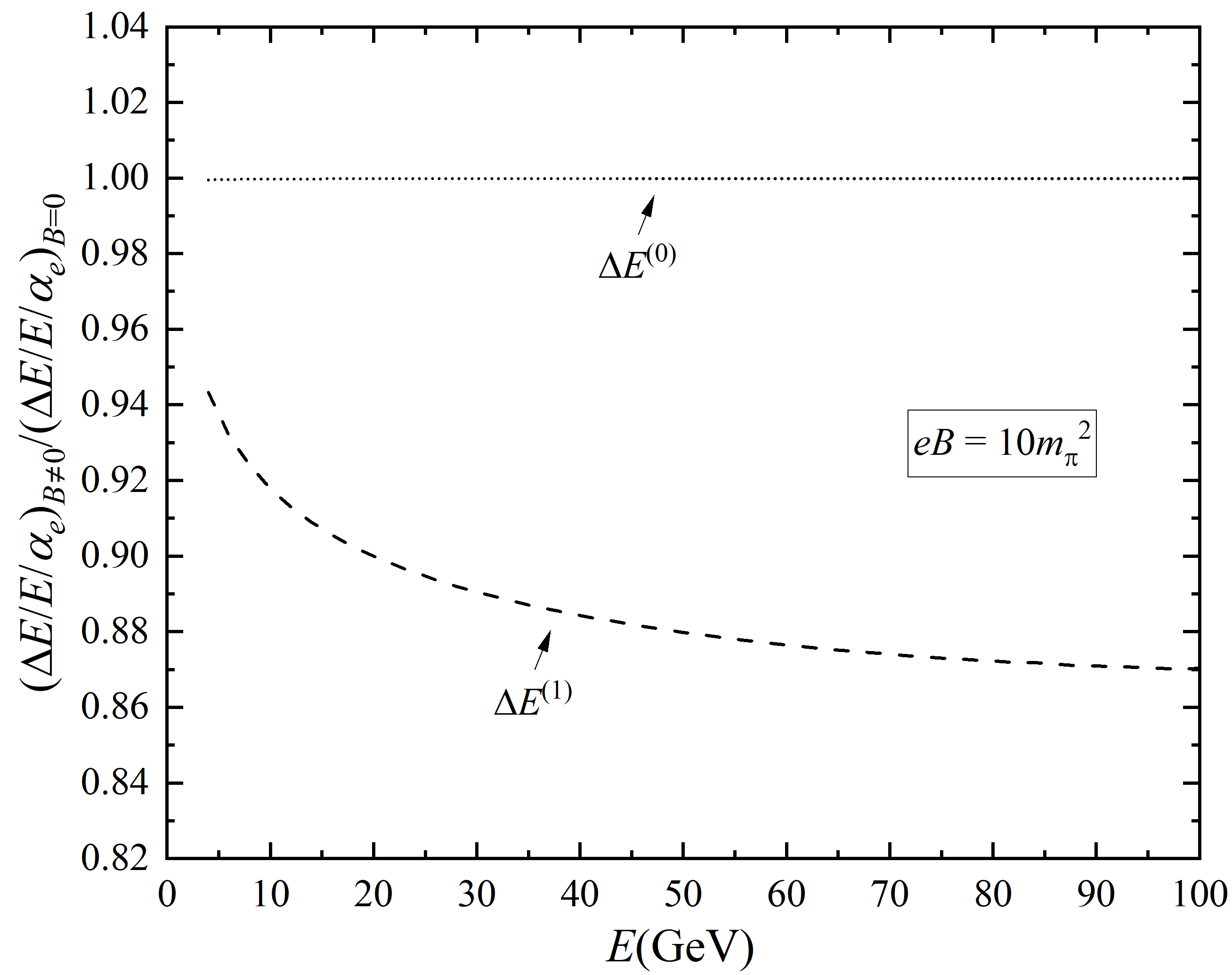}
	\caption{\label{fig:9}Ratios of the jet fractional electromagnetic energy loss as a function of jet initial energy $E$ with magnetic field to without magnetic field. The upper curve corresponds to the zeroth order in opacity, and the lower curve to the first order in opacity.}
\end{figure}
\begin{figure}[htbp]
	\centering
	\includegraphics[width=3.0in,height=2.34in]{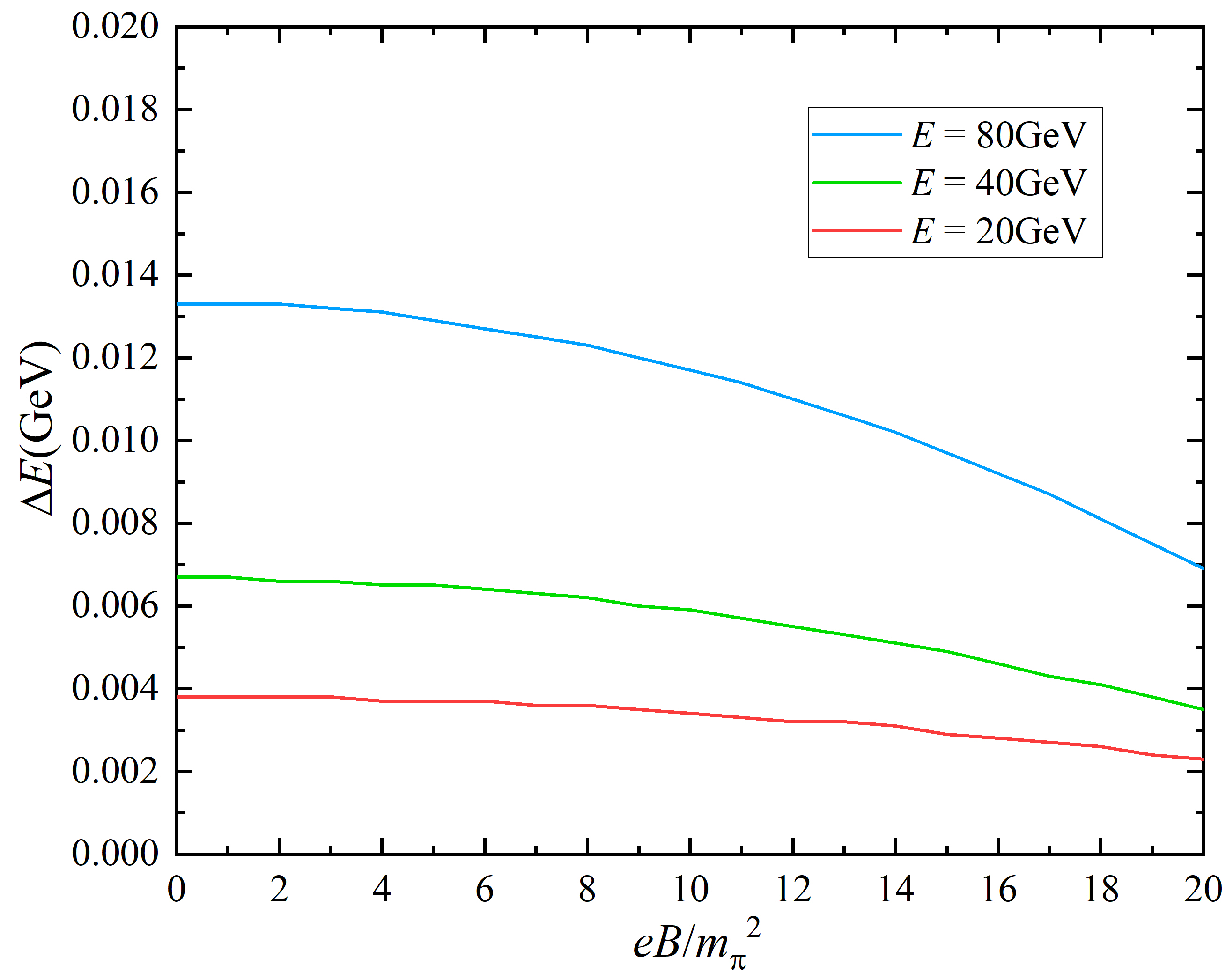}
	\caption{\label{fig:10}Jet electromagnetic energy loss as a function of the background magnetic field strength $B$ for different jet initial energies $E$.}
\end{figure}

In the Fig.~\ref{fig:7}, we show the comparison of jet fractional electromagnetic energy loss as a function of the jet initial energy with different strengths of background magnetic field. It is evident that the jet fractional energy loss due to electromagnetic radiation increases with the jet initial energy rises. This is in line with the analysis of the photon yield curve. Furthermore, the presence of the background magnetic field reduces the electromagnetic energy loss ratio of the jet.

Fig.~\ref{fig:8} further shows the ratios of jet fractional electromagnetic energy loss with different background magnetic fields and without magnetic field. We find that the background magnetic field moderately reduces the jet energy loss because of electromagnetic radiation. Specifically, when the magnetic field strength reaches $5m_{\pi}^{2}$ and $20m_{\pi}^{2}$, the jet energy loss decreases by approximately 1\% and 4\%, respectively, compared to the case with no magnetic field. This indicates that a stronger magnetic field leads to a greater reduction in the jet energy loss ratio. The trend aligns with the above conclusion that fewer photons are radiated, thereby carrying away less energy from the jet. Our result is consistent with the findings in Ref.~\cite{Dey:2023lco}, despite the different theoretical frameworks employed. Furthermore, as mentioned earlier, the higher-energy jets are less affected by magnetic field.

Fig.~\ref{fig:9} shows two ratios of fractional electromagnetic energy loss of a quark propagating in QGP medium with and without a magnetic field. $\Delta E^{(0)}$ denotes the zeroth order in opacity contribution, and $\Delta E^{(1)}$ denotes the first order in opacity contribution to the energy loss.
The results indicate that the magnetic suppression of electromagnetic energy loss mainly affects the photons radiated from jet–medium interactions, rather than the jet spontaneous emission. We propose that the correction terms introduced by the magnetic field complicate the LPM effect associated with multiple scattering, thereby enhancing the interference between scattering amplitudes and suppressing photon radiation.

Finally, the total energy loss for three different jet initial energies is plotted as the function of the background magnetic field strength in the Fig.~\ref{fig:10}.

Moreover, we have verified that increasing either the screening mass or the medium length (and thus the opacity) enhances both the photon yield and the corresponding jet energy loss, regardless of the presence of a magnetic field. The former stems from a shorter effective interaction range associated with a larger screening mass, which implies a higher effective density of scattering centers within the medium and thus stronger medium-induced radiation \cite{Baier:1998kq}. The latter, follows from the increased average number of scattering events along a longer propagation path, which amplifies the induced radiation—a trend well established in the literature \cite{Gyulassy:2000fs, Zhang:2010hiv}. Our numerical results remain consistent with these theoretical expectations across the explored range of parameters, confirming that the observed behavior is both physically sound and numerically robust. By fixing these magnetic-field-independent parameters, the influence of the magnetic effect on the photon radiation yield and the corresponding jet energy loss can be clearly demonstrated through the comparison of our theoretical calculations and potential experimental data.

\section{\label{sect:5}Summary}

In this study, the GLV opacity expansion is employed to investigate medium-induced photon radiation from a quark jet traversing a QGP in the presence of an background magnetic field in relativistic heavy-ion collisions. For theoretical tractability, the analysis is restricted to the case of a high-energy quark jet propagating parallel to the direction of the magnetic field.

Our numerical results indicate that the presence of a magnetic field leads to a slight overall suppression of medium-induced photon radiation from a quark jet. As the photon yield decreases, the corresponding electromagnetic radiative energy loss of the jet traversing the QGP is also moderately reduced.
Furthermore, we discover that a stronger background magnetic field results in a smaller jet electromagnetic energy loss.
The corrections introduced by the magnetic field are found to modify the LPM effect associated with multiple scattering and enhance the destructive interference among successive scattering events. 
Compared to the case without background magnetic field, this enhanced interference further suppresses photon emission and consequently diminishes the jet's radiative energy loss.

To compare photon yields with and without a magnetic field in AA collisions, specific experimental conditions must be carefully defined. At a fixed collision energy, the QGP produced in central collisions attains higher bulk temperatures than that formed in peripheral collisions. On the other hand, a background magnetic field is generated in peripheral collisions but remains negligible in central ones. With advances in detector technology and the accumulation of extensive experimental data, it is now feasible to probe the effects of magnetic fields with improved precision. A practical strategy is to select two collision systems---differing in geometry, centrality, and beam energy---such that the resulting fireballs have comparable size and temperature. This configuration would produce two QGP media that differ primarily in the strength of the internal magnetic field, thereby enabling a direct investigation of its influence on photon production and jet quenching.\\

\begin{acknowledgments}
The authors would like to thank He-Xia Zhang for helpful discussions. This work is supported in by National Natural Science Foundation of China under Grants Nos. 12535010.
\end{acknowledgments}

\begin{widetext}
\appendix
\section{\label{appe:1}The calculations about $\mathcal{M}_1$ and $\mathcal{M}_2$}
Throughout this paper, we use the metric convention $\eta_{\mu \nu} = \mathrm{diag}(+1,-1,-1, -1)$ and work in natural units where $c = 1$. When calculating the single rescattering diagrams, we need to handle such expression,
\begin{align}
	\mathcal{M}_{1,0} &= J(p+k)e^{i(p+k)\cdot x_0}(-i)g_e 2(E-\omega)\frac{2\bm{\epsilon}_{\perp} \cdot \bm{k}_{\perp}}{x} 
		\int \frac{\mathrm{d}^2 \bm{q}_{1\perp}}{(2\pi)^2}e^{-i\bm{q}_{1\perp}\cdot \bm{b}_{1\perp}}\nonumber\\
	&\quad \times \int \frac{\mathrm{d}q_{1z}}{2\pi} v(q_{1z},\bm{q}_{1\perp})\Delta(p+k-q_1)\Delta(p-q_1)e^{-iq_{1z}(z_1-z_0)} a_1T_{a_1},
\end{align}
where $\Delta$ is the quark jet propagator in magnetic field in weak field approximation. Isolate the integral over $q_{1z}$ and write it as
\begin{align}
	I_1 &=\int \frac{\mathrm{d}q_{1z}}{2\pi} v(q_{1z},\bm{q}_{1\perp}) \Delta(p+k-q_1)\Delta(p-q_1)e^{-iq_{1z}(z_1-z_0)}\nonumber\\
	&= \int \frac{\mathrm{d}q_{1z}}{2\pi} v(q_{1z},\bm{q}_{1\perp})
		\bigg\{\frac{1}{(p-q_1)^2-m^2+i\epsilon}-\frac{(qB)^2}{[(p-q_1)^2-m^2+i\epsilon]^3}\bigg\}\nonumber\\
	&\quad \times \bigg\{\frac{1}{(p+k-q_1)^2-m^2+i\epsilon}-\frac{(qB)^2}{[(p+k-q_1)^2-m^2+i\epsilon]^3}\bigg\} e^{-iq_{1z}(z_1-z_0)}\nonumber\\
	&\approx \int \frac{\mathrm{d}q_{1z}}{2\pi} v(q_{1z},\bm{q}_{1\perp})
		\bigg\{ \frac{1}{(p-q_1)^2-m^2+i\epsilon} \frac{1}{(p+k-q_1)^2-m^2+i\epsilon}\nonumber\\
	&\quad -\frac{1}{(p-q_1)^2-m^2+i\epsilon} \frac{(qB)^2}{[(p+k-q_1)^2-m^2+i\epsilon]^3}\nonumber\\
	&\quad - \frac{(qB)^2}{[(p-q_1)^2-m^2+i\epsilon]^3} \frac{1}{(p+k-q_1)^2-m^2+i\epsilon} \bigg\}e^{-iq_{1z}(z_1-z_0)},
\end{align}
where we have ignored the higher order term $\mathcal{O}[(qB)^4]$. We close the contour in lower half plane in the complex $q_{1z}$ plane since $z_1 > z_0$. The two propagators have two poles in the lower $q_{1z}$ plane, which are approximately located at $-i\epsilon$ and $-\omega_0 - i\epsilon$.
Here we neglect the exponentially suppressed pole at $-i\mu_1$ due to the assumption $z_1-z_0\gg 1/\mu$. After applying the residue theorem and some simplifications, the integration result is
\begin{align}
	I_1 \approx -iv(0,\bm{q}_{1\perp})\bigg[ \frac{1}{4E(E-\omega)\omega_0} + \frac{(qB)^2}{16E^3(E-\omega)\omega_0^3} + \frac{(qB)^2}{16E(E-\omega)^3\omega_0^3}\bigg] \big[1-e^{i\omega_0(z_1-z_0)}\big].
\end{align}
Substituting $I_1$ and $\omega_0$ into the original formula, we obtain
\begin{align}
	\mathcal{M}_{1,0} =-iJ(p+k)e^{i(p+k)\cdot x_0} \int \frac{\mathrm{d}^2 \bm{q}_{1\perp}}{(2\pi)^2} v(0,\bm{q}_{1\perp})
		e^{-i\bm{q}_{1\perp} \cdot \bm{b}_{1\perp}} a_1T_{a_1}\mathcal{R}_{1,0},
\end{align}
where
\begin{align}
	\mathcal{R}_{1,0} = -2ig_e(1-x)\frac{\bm{\epsilon}_{\perp} \cdot \bm{k}_{\perp}}{\bm{k}_{\perp}^{2} + x^2m^2} 
    \left[1 + \frac{(qB)^2x^2}{(\bm{k}_{\perp}^{2} + x^2m^2)^2} 
	+ \frac{(qB)^2x^2(1-x)^2}{(\bm{k}_{\perp}^{2} + x^2m^2)^2}\right] \big[1-e^{i\omega_0(z_1-z_0)}\big].
\end{align}
The calculation about $\mathcal{M}_{1,1}$ can be performed using the same method
\begin{align}
	\mathcal{R}_{1,1} = -2ig_e(1-x)\frac{\bm{\epsilon}_{\perp} \cdot (\bm{k}_{\perp}-x\bm{q}_{1\perp})}{(\bm{k}_{\perp}-x\bm{q}_{1\perp})^2 + x^2m^2} \bigg\{1-\frac{(qB)^2x^2(1-x)^2}{[(\bm{k}_{\perp}-x\bm{q}_{1\perp})^2 + x^2m^2]^2} \bigg\}e^{i\omega_0(z_1-z_0)}.
\end{align}

We now turn to the computation of the double Born rescattering diagrams and consider expressions of the form:
\begin{align}
	\mathcal{M}_{2,0} &\approx J(p+k)e^{i(p+k)\cdot x_0}(-i)g_e 4(E-\omega)^2\frac{2\bm{\epsilon}_{\perp} \cdot \bm{k}_{\perp}}{x}
		\int \frac{\mathrm{d}^2 \bm{q}_{1\perp}}{(2\pi)^2}e^{-i\bm{q}_{1\perp}\cdot \bm{b}_{1\perp}}
		\int \frac{\mathrm{d}^2 \bm{q}_{2\perp}}{(2\pi)^2}e^{-i\bm{q}_{2\perp}\cdot \bm{b}_{2\perp}}\nonumber\\
	&\quad \times \int \frac{\mathrm{d}q_{2z}}{2\pi} v(q_{2z},\bm{q}_{2\perp})\Delta(p-q_2) e^{-iq_{2z}(z_2-z_1)}\nonumber\\
	&\quad \times \int \frac{\mathrm{d}q_{1z}}{2\pi} v(q_{1z},\bm{q}_{1\perp}) \Delta(p+k-q_1-q_2) \Delta(p-q_1-q_2) e^{-i(q_{1z}+q_{2z})(z_1-z_0)} a_2a_1T_{a_2}T_{a_1}.
\end{align}
Isolate the integral over $q_{1z}$ and write it as
\begin{align}
	I_2 &=\int \frac{\mathrm{d}q_{1z}}{2\pi} v(q_{1z},\bm{q}_{1\perp}) \Delta(p+k-q_1-q_2)\Delta(p-q_1-q_2)e^{-i(q_{1z}+q_{2z})(z_1-z_0)}\nonumber\\
	&= \int \frac{\mathrm{d}q_{1z}}{2\pi} v(q_{1z},\bm{q}_{1\perp})
		\bigg\{\frac{1}{(p+k-q_1-q_2)^2-m^2+i\epsilon}-\frac{(qB)^2}{[(p+k-q_1-q_2)^2-m^2+i\epsilon]^3}\bigg\}\nonumber\\
	&\quad \times \bigg\{\frac{1}{(p-q_1-q_2)^2-m^2+i\epsilon}-\frac{(qB)^2}{[(p-q_1-q_2)^2-m^2+i\epsilon]^3}\bigg\}
		e^{-i(q_{1z}+q_{2z})(z_1-z_0)}\nonumber\\
	&\approx \int \frac{\mathrm{d}q_{1z}}{2\pi} v(q_{1z},\bm{q}_{1\perp})
		\bigg\{ \frac{1}{(p+k-q_1-q_2)^2-m^2+i\epsilon} \frac{1}{(p+q_1-q_2)^2-m^2+i\epsilon}\nonumber\\
	&\quad -\frac{1}{(p+k-q_1-q_2)^2-m^2+i\epsilon} \frac{(qB)^2}{[(p-q_1-q_2)^2-m^2+i\epsilon]^3}\nonumber\\
	&\quad - \frac{(qB)^2}{[(p+k-q_1-q_2)^2-m^2+i\epsilon]^3} \frac{1}{(p-q_1-q_2)^2-m^2+i\epsilon} \bigg\}e^{-i(q_{1z}+q_{2z})(z_1-z_0)},
\end{align}
where we have ignored the higher order term about $(qB)^4$ again. The two propagators have two poles in the lower $q_{1z}$ plane, which are approximately located at $-q_{2z} -\omega_0 -i\epsilon$ and $-q_{2z} - i\epsilon$.
We integrated over $q_{1z}$ by closing the contour in lower half plane and neglected the pole at $-i\mu_1$ again and set $x_1=x_2=x_j$ here for the contact limit. After applying the residue theorem and some simplifications, the integration result is
\begin{align}
	I_2 \approx -iv(-q_{2z},\bm{q}_{1\perp})\bigg[ \frac{1}{4E(E-\omega)\omega_0} + \frac{(qB)^2}{16E^3(E-\omega)\omega_0^3} + \frac{(qB)^2}{16E(E-\omega)^3\omega_0^3}\bigg] \big[1-e^{i\omega_0(z_1-z_0)}\big].
\end{align}
Unlike single rescattering, we still have to integrate $q_{2z}$, that is
\begin{align}
	I_3 &=\int \frac{\mathrm{d}q_{2z}}{2\pi} v(q_{2z},\bm{q}_{2\perp})v(-q_{2z},\bm{q}_{1\perp})\Delta(p-q_2)e^{-iq_{2z}(z_2-z_1)}\nonumber\\
	&= \int \frac{\mathrm{d}q_{2z}}{2\pi} v(q_{2z},\bm{q}_{2\perp})v(-q_{2z},\bm{q}_{1\perp}) 
        \bigg\{ \frac{1}{(p-q_2)^2-m^2+i\epsilon} -\frac{(qB)^2}{[(p-q_2)^2-m^2+i\epsilon]^3} \bigg\}e^{-iq_{2z}(z_2-z_1)}.
\end{align}
Close the contour in lower half plane and integrate over $q_{2z}$. 
Note that besides the pole at $-i\epsilon$ comes from the propagator, the poles at $-i\mu_1$ and $-i\mu_2$ also contribute to the integral as there is no exponentially suppressed factor. The integration result is
\begin{align}
	I_3 \approx -iv(0,\bm{q}_{1\perp})v(0,\bm{q}_{2\perp}) 
        \bigg[ \frac{1}{4(E-\omega)} - \frac{(qB)^2}{16(E-\omega)^3}\frac{\mu_1^2+\mu_2^2}{\mu_1^2\mu_2^2}\bigg].
\end{align}
Substituting $I_2$, $I_3$ and $\omega_0$ into the original formula, we obtain
\begin{align}
	\mathcal{M}_{2,0} = -iJ(p+k)e^{i(p+k)\cdot x_0} \int \frac{\mathrm{d}^2 \bm{q}_{1\perp}}{(2\pi)^2} v(0,\bm{q}_{1\perp}) \int \frac{\mathrm{d}^2 \bm{q}_{2\perp}}{(2\pi)^2} v(0,\bm{q}_{2\perp}) e^{-i(\bm{q}_{1\perp}+\bm{q}_{2\perp})\cdot \bm{b}_{j\perp}} a_2a_1T_{a_2}T_{a_1}\mathcal{R}_{2,0},
\end{align}
where
\begin{align}
	\mathcal{R}_{2,0} = -g_e(1-x)\frac{\bm{\epsilon}_{\perp} \cdot \bm{k}_{\perp}}{\bm{k}_{\perp}^2 + x^2m^2}
		\bigg[1 + \frac{(qB)^2x^2}{(\bm{k}_{\perp}^{2} + x^2m^2)^2} 
		+ \frac{(qB)^2x^2(1-x)^2}{(\bm{k}_{\perp}^{2} + x^2m^2)^2}\bigg]\big[1-e^{i\omega_0(z_j-z_0)}\big].
\end{align}
The calculation about $\mathcal{M}_{2,2}$ can be performed using the same method and
\begin{align}
    \mathcal{R}_{2,2} &= -g_e(1-x)\frac{\bm{\epsilon}_{\perp}\cdot(\bm{k}_{\perp}-x\bm{q}_{1\perp}-x\bm{q}_{2\perp})}
    	{(\bm{k}_{\perp}-x\bm{q}_{1\perp}-x\bm{q}_{2\perp})^2 + x^2m^2}
    	\bigg\{1 - \frac{(qB)^2x^2(1-x)^2}{[(\bm{k}_{\perp} - x\bm{q}_{1\perp}-x\bm{q}_{2\perp})^2 + x^2m^2]^2}\bigg\} e^{i\omega_0(z_j-z_0)}.
\end{align}

\section{\label{appe:2}The reorganization of scattering amplitude}

The radiation amplitudes of self-quenching, single rescattering and double Born rescattering are respectively given as follows
\begin{align}
	\mathcal{R}_0 &= 2g_e(1-x)\bm{\epsilon}_{\perp} \cdot \left[\bm{H} - x^2(1-x)^2\bm{H}^B\right],\nonumber\\
	\mathcal{R}_{1} &= -2ig_e(1-x)\bm{\epsilon}_{\perp} \cdot \big\{ \bm{H} + x^2\bm{H}^B + x^2(1-x)^2\bm{H}^B\nonumber\\
        &\quad - \big[\bm{H} + x^2\bm{H}^B + x^2(1-x)^2\bm{H}^B - \bm{C}_1 + x^2(1-x)^2\bm{C}_1^B \big] e^{i\omega_0(z_1-z_0)}\big\},\nonumber\\
	\mathcal{R}_{2} &= -g_e(1-x)\bm{\epsilon}_{\perp} \cdot \big\{ \bm{H} + x^2\bm{H}^B + x^2(1-x)^2\bm{H}^B - \left[x^2\bm{H}^B + 2x^2(1-x)^2\bm{H}^B\right] e^{i\omega_0(z_j-z_0)} \big\}.
\end{align}
Perform the ensemble average over $\mathcal{R}_{1}$,
\begin{align}
	{\mathrm{Tr}}\left \langle |\mathcal{M}_1|^2 \right \rangle = d_Rd_T|J(p)|^2\frac{L}{\lambda}
		\int \mathrm{d}^2 \bm{q}_{\perp} |\bar{v}(0,\bm{q}_{\perp})|^2 \int \mathrm{d}z_0 \int \mathrm{d}z \rho(z_0,z)|\mathcal{R}_1|^2,
\end{align}
where
\begin{align}
	|\mathcal{R}_1|^2 &= 4g_e^2(1-x)^2 \big\{ 2\bm{H}^2 + 4x^2\bm{H}\cdot\bm{H}^B + 4x^2(1-x)^2\bm{H}\cdot\bm{H}^B - 2\bm{H}\cdot\bm{C}_1 + 2x^2(1-x)^2\bm{H}\cdot\bm{C}_1^B\nonumber\\
    &\quad - 2x^2\bm{H}^B\cdot\bm{C}_1 - 2x^2(1-x)^2\bm{H}^B\cdot\bm{C}_1 + \bm{C}_1^2 - 2x^2(1-x)^2\bm{C}_1\cdot\bm{C}_1^B\nonumber\\
    &\quad - \cos[\omega_0(z-z_0)]\big[2\bm{H}^2 + 4x^2\bm{H}\cdot\bm{H}^B + 4x^2(1-x)^2\bm{H}\cdot\bm{H}^B\nonumber\\
    &\quad + 2x^2(1-x)^2\bm{H}\cdot\bm{C}_1^B- 2\bm{H}\cdot\bm{C}_1 - 2x^2\bm{H}^B\cdot\bm{C}_1 - 2x^2(1-x)^2\bm{H}^B\cdot\bm{C}_1\big]\big\}.
\end{align}
Subsequently, perform the ensemble average over $2{\mathrm{Re}}(\mathcal{M}_2 \mathcal{M}_0^*)$,
\begin{align}
	{\mathrm{Tr}}\left \langle 2{\mathrm{Re}}(\mathcal{M}_2 \mathcal{M}_0^*) \right \rangle
	= d_Rd_T|J(p)|^2 \frac{L}{\lambda} \int \mathrm{d}^2 \bm{q}_{\perp} |\bar{v}(0,\bm{q}_{\perp})|^2
	\int \mathrm{d}z_0 \int \mathrm{d}z \rho(z_0,z)2{\mathrm{Re}}(\mathcal{R}_2 \mathcal{R}_0^*),
\end{align}
where
\begin{align}
	2{\mathrm{Re}}(\mathcal{R}_2 \mathcal{R}_0^*) = -4g_e^2(1-x)^2 \big\{ \bm{H}^2 + x^2\bm{H}\cdot\bm{H}^B 
	- \cos[\omega_0(z-z_0)]\big[ x^2\bm{H}\cdot\bm{H}^B + 2x^2(1-x)^2\bm{H}\cdot\bm{H}^B \big]\big\}.
\end{align}
Finally, we obtain the expression in first order in opacity,
\begin{align}
	{\mathrm{Tr}}\left \langle |\mathcal{M}_1|^2 + 2{\mathrm{Re}}(\mathcal{M}_2 \mathcal{M}_0^*) \right \rangle
	&= d_Rd_T|J(p)|^2\frac{L}{\lambda}\int \frac{\mu^2\mathrm{d}^2\bm{q}_{\perp}}{\pi(\bm{q}_{\perp}^2 + \mu^2)^2}
		\int \mathrm{d}z_0 \int \mathrm{d}z \rho(z_0,z) \big[|\mathcal{R}_1|^2 + 2{\mathrm{Re}}(\mathcal{R}_2 \mathcal{R}_0^*)\big]\nonumber\\
	&= 4g_e^2 d_Rd_T|J(p)|^2(1-x)^2\frac{L}{\lambda}\int \mathrm{d}^2 \bm{q}_{\perp} |\bar{v}(0,\bm{q}_{\perp})|^2
        \int \mathrm{d}z_0 \int \mathrm{d}z \rho(z_0,z)\nonumber\\
	&\quad \times \big\{(\bm{H} - \bm{C}_1)^2 + x^2(3 + 4(1-x)^2)\bm{H}\cdot\bm{H}^B\nonumber\\
	&\quad + 2x^2(1-x)^2(\bm{H}\cdot\bm{C}_1^B - \bm{C}_1\cdot\bm{C}_1^B) - 2x^2(1 + (1-x)^2)\bm{H}^B\cdot\bm{C}_1\nonumber\\
	&\quad -\cos [\omega_0(z-z_0)] \big[2(\bm{H}^2-\bm{H}\cdot\bm{C}_1) + x^2(3 + 2(1-x)^2)\bm{H}\cdot\bm{H}^B\nonumber\\
	&\quad + 2x^2(1-x)^2\bm{H}\cdot\bm{C}_1^B - 2x^2(1 + (1-x)^2)\bm{H}^B\cdot\bm{C}_1 \big] \big\}.
\end{align}
\end{widetext}

\nocite{*}

\bibliography{apssamp}

\end{document}